\documentclass[journal]{IEEEtran}

\pdfoutput=1
\usepackage{amsmath}
\usepackage{amssymb}
\usepackage{amscd}
\usepackage{textcomp}
\usepackage{fancyvrb}
\usepackage{wasysym}
\usepackage{cite} 
 
\usepackage{graphicx}

\usepackage{listings}    

\newcommand{\be}{\begin{equation}}
\newcommand{\ee}{\end{equation}}
\newcommand{\bea}{\begin{eqnarray}}
\newcommand{\eea}{\end{eqnarray}}
\newcommand{\bal}{\begin{align}}
\newcommand{\eal}{\end{align}}
\newcommand{\nn}{\nonumber}
\newcommand{\eye}{\mbox{$\mbox{1}\!\mbox{l}\;$}}

\renewcommand{\vec}[1]{\boldsymbol{#1}}

\renewcommand{\P}{\mathcal{P}}
\renewcommand{\phi}{\varphi}

\begin{document}
\title{From State Estimation to Network Reconstruction}

\author{
Farnaz~Basiri,%
\thanks{F.~Basiri is with the Forschungszentrum J\"ulich, Institute for Energy and Climate Research -
	Systems Analysis and Technology Evaluation (IEK-STE),  52428 J\"ulich.}
Jose Casadiego,%
\thanks{J.~Casadiego and M.~Timme are with the Max Planck Institute for Dynamics and Self-Organization (MPIDS),  37077 G\"ottingen, Germany.}
Marc Timme, 
Dirk~Witthaut%
\thanks{D.~Witthaut is with the Forschungszentrum J\"ulich, Institute for Energy and Climate Research -
	Systems Analysis and Technology Evaluation (IEK-STE),  52428 J\"ulich, Germany
and the Institute for Theoretical Physics, University of Cologne, 
50937 K\"oln, Germany.}
}


\markboth{Journal of \LaTeX\ Class Files,~Vol.~13, No.~9, September~2014}%
{Shell \MakeLowercase{\textit{et al.}}: Bare Demo of IEEEtran.cls for Journals}

\maketitle

\begin{abstract}
We develop methods to efficiently reconstruct the topology and line parameters of a power grid from the measurement of nodal variables.  We propose two compressed sensing algorithms that minimize the amount of necessary measurement resources by exploiting network sparsity, symmetry of connections and potential prior knowledge about the connectivity. The algorithms are reciprocal to established state estimation methods, where nodal variables are estimated from few measurements given the network structure. Hence, they enable an advanced grid monitoring where both state and structure of a grid are subject to uncertainties or missing information. 
\end{abstract}


\IEEEpeerreviewmaketitle


\section{Introduction}

The secure and reliable operation of complex power grids requires a precise knowledge of the grid topology and the state of connected generation and transmission elements. A central tool is `state estimation', i.e the estimation of the current state of the nodes from few, possibly noisy measurements and the grid data. Various methods and algorithms have been developed in this field, see e.g.~\cite{Meli01} for a review. A prime example is the estimation of the voltage phase angles from power flow measurements (see e.g.~\cite{Wood14} for basic textbook examples).
State estimation algorithms typically assume that the network structure, i.e. the location and parameters of transmission lines and transformers, are known. However, parts of the information might be lacking or uncertain, for example for contingencies or attacks, such that the question arises: Is it possible to reconstruct the network structure from local measurements only? 

In this article we develop two algorithms that enable the faithful reconstruction of the network structure or parts of it from the measurement of nodal variables. Special attention is paid to the efficiency of these methods: We show how to reduce the number of measurements using methods from compressed sensing exploiting the structural properties of power grids. We here focus on the DC approximation to formulate the fundamental ideas of network reconstruction most clearly. 

The presented algorithms can find several applications in the monitoring and operation of (smart) power grids. While it is unlikely that the entire network structure is unknown, situations regularly occur where parts of the information is lacking. Manual switches still exist in many distribution grids, transmission lines can get lost in contingency cases and targeted attacks can cause entire regions of a grid to collapse. We present an example where the connectivity in two regions of the grid is a priori unknown, but can be reconstructed from a single snapshot of the nodal variables. 
In future power grids, the network's structure will become even more variable: The transmission of high-voltage direct current (HVDC) lines in hybrid power grids can be actively controlled and the effective line parameters can be regulated with flexible alternating current transmission system (FACTS) devices. Then it becomes essential to include lacking information or uncertainties of the grid topologies in any grid monitoring tool. The methods presented in this article shall contribute to the development of such advanced monitoring algorithms \cite{Yuan11,Gu12,Bolo14}.

\section{The DC approximation}

The DC approximation describes the flow of real power flow in AC power grids in a linearized way. It is derived from the full nonlinear load flow equations under the following additional simplifying assumptions  \cite{Grai94,Wood14,Purch05,Hert06}:
\begin{itemize}
\item 	All bus voltage magnitudes $|V_s|$ are close to 1 p.u. and approximated to be exactly $|V_s| = 1$ p.u.
\item Voltage angle differences across branches are small, such that $\sin(\phi_{s} - \phi_r) \approx \phi_{s} - \phi_r$ 
   and $\cos(\phi_{s} - \phi_r) \approx 1$. 
\item	Branches can be considered lossless. In particular, branch resistances and charging capacitances are negligible. 
\end{itemize}
The approximate real power flow from node $s$ to node $r$ over a transmission line with reactance
$X_{sr}$ is then given by
\be
   P_{sr} =\frac{|V_s | |V_r |}{X_{sr}} \sin(\phi_{s} - \phi_r)
      \approx X_{sr}^{-1} (\phi_{s} - \phi_r)
\ee
where $\phi_s$ is the phase angle at bus $s$. 
The real power balance at the node $s$ reads
\be
   P_s = \sum_{r=1}^N P_{sr} =  \sum_{r=1}^N X_{sr}^{-1} (\phi_{s} - \phi_r).
\ee

For notational convenience we summarize these equations in matrix form. We define the vector of all voltage angles and power injections, 
\begin{align*} 
   \vec \phi &= (\phi_1,\ldots,\phi_N)^T \in \mathbb{R}^N,
   \vec P &= (P_1,\ldots,P_N)^T \in \mathbb{R}^N,
\end{align*}
where $N$ is the number of buses or nodes in the grid and the superscript `$T$' denotes the transpose of a vector or matrix. The DC approximation then yields the condition
\be
    \vec B \vec \phi  = \vec P,
    \label{eq:dcapprox}
\ee
where the nodal susceptance matrix $\vec B \in \mathbb{R}^{N \times N}$ 
has elements
\begin{equation}
  B_{nk} = \left\{ 
   \begin{array}{lll}
   \displaystyle\sum \nolimits_{j=1}^{N} X_{n j}^{-1} &  \mbox{if} & k = n; \\ [2mm]
     - X_{nk}^{-1} & \mbox{if} & k \neq n.
   \end{array} \right.
   \label{eq:defB}
\end{equation}
The matrix $\vec B$ is a  Laplacian matrix, which has one zero eigenvalue 
with eigenvector $(1,1,\ldots,1)^T$ \cite{Newm10}. This eigenvector 
represents a global shift of all voltage angles without physical 
significance.
To avoid this complication, one often fixes the voltage angle at a reference bus and excludes this bus from the calculation, i.e. the corresponding row and column will be missing in $\vec B$.
We finally note that mathematically equivalent models of flow are used 
to describe hydraulic networks \cite{Hwan96} or vascular networks of plants 
\cite{Kati10}.

\section{State Estimation}
\label{sec:stateest}

State estimation is a mathematical procedure by which the state of an electric power system is extracted from a set of measurements \cite{Meli01}. In general, any measurement can be expressed as a function of the system state. Let 
\be
   z_i=h_i (\vec x)                                     
\ee   
denote a measured quantity, where $\vec x$ is the system state and $h_i$ is a function specific to the measured quantity $z_i$. Assume that $M$ measurements are taken. Then, all measurements can be written in compact form
\be
    \vec z = \vec h( \vec x)
    \label{eq:stateest1}
\ee
where 
\begin{align*}
& \vec x \in \mathbb{R}^{N \times 1} \;    \mbox{is the system state,} \\
& \vec z \in \mathbb{R}^{M \times 1} \;     \mbox{is a vector of measured quantities and} \\
& \vec h \; \mbox{is a vector function, i.e. a mapping} \, \mathbb{R}^N \rightarrow \mathbb{R}^M
\end{align*}
Typically more measurements than the number of state variables to be determined are taken, i.e. $M>N$, such that equation (\ref{eq:stateest1}) is overdetermined. It is then solved in a least-squares fashion, i.e. one calculates $\vec x$ such that $\| \vec z - \vec h( \vec x)  \|_2^2$  is minimized. For underdetermined equations (\ref{eq:stateest1}) typically many solutions exist. 

It is often helpful to work with a simplified DC approximation model for the measurement equations in analyzing the inherent limitations of various methods related solely to the measurement configuration \cite{Abur04}. The state of the grid is described by the voltage phase angles $\vec x = \vec \phi$ and shall be estimated from measurements. As explicated above, the real power flow measured from bus $s$ to $r$ can be approximated by
\be
      P_{sr} = X_{sr}^{-1} (\phi_{s} - \phi_r) + e_{sr}
\ee
where $\phi_s$ is the phase angle at bus $s$ and $e_{sr}$ is a measurement error. Similarly, a power injection measurement 
\be
        P_s = \sum_{r=1}^N P_{sr} + e'_s 
\ee
at a given bus $s$ can be expressed as a sum of flows along incident branches to that bus. The measurement vector $\vec z$ thus consists of a subset of the power flows $P_{sr}$ and power injections $P_s$. In the DC approximation the vector function $\vec h$ are all linear, which strongly simplifies the problem such that it is especially suitable to introduce new concepts such as network reconstruction.

\section{Network Reconstruction}

Let us come back to the defining equation for the DC approximation
\be
    \vec B \vec \phi = \vec P.
\ee
In state estimation, we assume that the nodal phase angles $\vec \phi$ are difficult to measure and shall be estimated from the knowledge of the two other quantities in this equation, $\vec B$ and $\vec P$. But we can also reverse the problem and ask what happens if we don't know the nodal susceptance matrix $\vec B$. Can we efficiently reconstruct its entries from measurements of the two remaining quantities $\vec P$ and $\vec \phi$? 
This problem may appear hypothetical at a first glance, as the network structure and parameters are generally known. But there are situations where at least parts of the network structure are unknown: Switches can be open or closed, transmission lines can undergo failures, or plans may be inaccurate. Even more, in times of war or terrorism we may be extremely unsure about the physical integrity of parts of the grid. Can we reconstruct the structure of the grid from measurements only?

In the following we develop a mathematical theory of network reconstruction. To introduce the method, we start from the hypothetical situation that $\vec B$ is completely unknown, whereas $\vec P$ and $\vec \phi$ can be measured perfectly. We then study more realistic cases, where only parts of the grid topology are unknown. 

We first notice that a single measurement of $\vec P$ and $\vec \phi$ will obviously not be enough to reconstruct the entire grid topology. However, we may repeat our measurement at different times $t_1,t_2,\ldots,t_M$ such that we obtain a large number of conditions on the entries of $\vec B$ of the form  
\be
    \sum_k B_{rk} \phi_k(t_m) =  P_r(t_m) 
    \label{eq:BPPconditions}
\ee    
which hold for all rows $r = 1,\ldots,N$ and all measurement time steps $m=1,\ldots, M$. Furthermore we know that the row-sums of the nodal susceptance matrix (\ref{eq:defB}) vanish,
\be
    \sum_k B_{rk}  = 0.
     \label{eq:rowsum}
\ee
In total we thus have $N \times (M+1)$ conditions which we can use to reconstruct the unknown entries of $\vec B$.

\subsection{Row-wise reconstruction} 
\label{sec:idea-rowwise}

In a first approach we aim to reconstruct the nodal admittance matrix row-by-row (cf.~\cite{Timme2007,Shandilya2011,Timme2014,Han2015}). To simplify notation we collect all measurement conditions (\ref{eq:BPPconditions}) and the condition (\ref{eq:rowsum}) for a given row $r$ and rewrite them in matrix form
\be
   \underbrace{\begin{pmatrix}
    \phi_1(t_1) & \cdots & \phi_N(t_1) \\
    \vdots &  & \vdots \\
    \phi_1(t_M) & \cdots & \phi_N(t_M) \\
    1 & \cdots & 1 \\ 
   \end{pmatrix}}_{=: \vec \Phi}
   \underbrace{\begin{pmatrix}
       B_{r1} \\ \vdots \\ B_{rN}
   \end{pmatrix}}_{=: \vec B_r} 
   =
   \underbrace{\begin{pmatrix}
     P_{r}(t_1) \\ \vdots \\ P_{r}(t_M) \\ 0
   \end{pmatrix}}_{=: \vec \P_r} 
   \label{eq:PhiBP1}
\ee
Here, $\vec B_r \in \mathbb{R}^{N\times 1}$ denotes the transpose of the $r$th row of the nodal admittance matrix $\vec B$, which we want to reconstruct. The matrix $\vec \Phi \in \mathbb{R}^{(M+1)\times N}$ summarizes all measurement results for the voltage phase angles at all time steps, while the vector $\vec \P_r \in \mathbb{R}^{(M+1)\times 1}$ summarizes the values of the power injection only for node $r$. The last row of the matrix equation represents the condition (\ref{eq:rowsum}) for the row sum.

Now we have to distinguish, whether equation (\ref{eq:PhiBP1}) is (over)determined or underdetermined. If the number of linearly independent equations is larger or equal to $N$ we generally have enough information to directly compute $\vec B_r$. In particular there is exactly one solution if
${\rm rank}(\bar {\vec \Phi} | \bar{ \vec \P_r} ) =  {\rm rank}(\bar {\vec \Phi}) = N$, which can be calculated using Gaussian elimination. Nevertheless, matrix inversion can be numerically ill-conditioned, especially for systems with large $N$ \cite{Stoer93}.
If the data includes some measurement noise, the system of equations (\ref{eq:PhiBP1}) can be overdetermined when measuring $M+1>N$ times such that it is solvable in a least squares fashion, i.e. we have to determine the vector $\vec B_r \in \mathbb{R}^{N\times 1}$ minimizing the 2-norm
\be
    \| \vec \Phi \vec B_r - \vec \P_r     \|_2^2 \, =
       \sum_{k=1}^{M+1}  (\vec \Phi \vec B_r - \vec \P_r)_k^2. 
\ee 
To efficiently reconstruct the grid topology, we want to rely on as few measurements as possible. If $M+1 < N$, we generally have an underdetermined set of equations which admits many possible solutions. Is it possible to obtain the correct solution also in this case? Do we have more information about $\vec B_r$ which we can exploit?
Indeed, we know that a power grid is typically very sparse -- a single substation is connected to only few other substations. Thus we choose the one solution to equation (\ref{eq:PhiBP1}) which minimizes the number of non-zero entries of $\vec B_r$. Unfortunately, the direct minimization is computationally hard in general. In 2006 Candes, Romberg, Tao and Donoho showed that an efficient reconstruction is nevertheless possible using a convex surrogate for sparsity: the 1-norm \cite{Dono06,Cand06}. Under weak conditions, the correct sparse solution can be calculated efficiently by minimizing the 1-norm 
\be
     \| \vec B_r \|_1 = \sum_{k=1}^N |\vec B_{rk}|
     \label{eq:Br-1norm}
\ee
subject to the constraint (\ref{eq:PhiBP1}). This problem can be mapped to a linear program which we explicate in the following section.

\subsection{Partial reconstruction}
\label{sec:idea-partial}

In applications we will rarely encounter the situation that the structure of a power grid is completely unknown. In a typical application we know the value of $B_{rs}$ of all transmission lines $(s,r)$ in one part of the grid and have to reconstruct only the remaining entries of the matrix $\vec B$ -- for instance we might want to monitor the position of manual switches at remote places of the grid.

To keep track of our knowledge about the network, we define the matrix $\vec K \in \mathbb{R}^{N\times N}$ with entries
\be
    K_{rs} := \left\{
    \begin{array}{l l}
    1 & \; \mbox{if the value of $B_{rs}$ is known for line $(r,s)$}  \\
    0 & \; \mbox{otherwise}.
    \end{array} \right.
\ee
We again proceed row by row. For the $r$th row we reduce equation (\ref{eq:PhiBP1}) to 
\be
    \vec \Phi_r^{\rm red} \vec B_r^{\rm red} = \vec \P_r^{\rm red},
    \label{eq:PhiBR-red}
\ee
where
\be
    \vec \P_r^{\rm red} = \vec \P_r - \sum_{c=1}^N  B_{rc} K_{rc} 
     \begin{pmatrix}
         \phi_c(t_1) \\  \vdots \\ \phi_c(t_M) \\  1 
   \end{pmatrix}
\ee
and
$\vec \Phi_r^{\rm red}$ is the submatrix of $\vec \Phi_r$ obtained by deleting all columns $c$ for which $K_{rc} = 1$ and $\vec B_r^{\rm red}$ is the submatrix of $\vec B_r$ obtained by deleting all rows $r'$ for which $K_{r,r'} = 1$.
The dimension of the reduced linear system of equations (\ref{eq:PhiBR-red}) is smaller than in the original problem (\ref{eq:PhiBP1}), such that the correct solution can typically be found from fewer measurements.

%

\subsection{Reconstruction from power flow measurements}

The reconstruction scheme introduced above requires the knowledge of the voltage phase angle at all nodes of the grid. This data can in principle be obtained using phasor measurement units (PMUs), but these are typically very expensive. The measurement of other quantities such as real power flows is typically much simpler.

Fortunately, we usually do not have to reconstruct the entire grid in most cases. In a typical application we know a lot about the grid and have to reconstruct only parts of the matrix $\vec B$. Then the measurement of power injections and real power flows of a certain subset of transmission lines can be enough to perform the reconstruction of the remaining data using the methods described above. Technically, this amounts to combining methods of classical state estimation and network reconstruction.

So assume that we can measure  the power injection $P_n$ for all nodes $n\in\{1,\ldots,N\}$ at the time steps $t_1,\ldots,t_M$. In addition we have knowledge about the transmission line parameters, in particular $X_{sr}$, and measure the real power flow
\be
    P_{sr}(t_m) =  X_{sr}^{-1} \big( \phi_r(t_m) - \phi_s(t_m) \big)   
\ee
for a subset $\mathcal{L}$ of all transmission lines at all time steps $t_1,\ldots,t_M$. If this set of equations is fully determined or even overdetermined we reconstruct the state vector $\vec \phi(t_m)$  as in the classical state estimation problem described in section \ref{sec:stateest}. As before the resulting estimates for the nodal phase angles for all time steps $t_1,\ldots,t_M$ are then summarized in the matrix $\vec \Phi$. In addition we have partial knowledge about the nodal susceptance matrix, in particular we already know the entries
\be
   B_{rs} = B_{sr}  = - X_{sr}^{-1}
\ee  
for all $(s,r) \in \mathcal{L}$. This information can then be used to reconstruct the remaining entries of $\vec B$ as described in the previous section.

\section{Implementation}

\subsection{Least-squares solution}

If we have many measurements or prior knowledge available, then the system of equations (\ref{eq:PhiBP1}) can be be overdetermined. In this case equation must be solved in a least squares fashion, i.e. we have to solve
\be
    \min_{\vec B_r} \| \vec \Phi \vec B_r - \vec \P_r     \|_2^2 \, ,
\ee 
which leads to \cite{Timme2014}
\be
   \vec B_r = (\vec \Phi^T \vec \Phi)^{-1} \, \vec \Phi^T \, \vec \P_r
\ee
provided that $\vec \Phi$ is full rank such that the inverse exists. Least-squares solutions are already implemented in many numerical solvers, for instance in the {\tt MATLAB} function {\tt mldivide}.

\subsection{Minimizing the 1-norm}
\label{sec:1norm}

The key to an efficient reconstruction of $\vec B$ from an underdetermined system of equations is the minimization of the 1-norm \cite{Dono06,Cand06}. This problem can be mapped to a linear program which can be solved efficiently, i.e.~the computation time scales at most polynomially with the input length.

So we consider the linear system of equations (\ref{eq:PhiBP1})  and assume that it is solvable and underdetermined. The solutions span an affine subspace of of $\mathbb{R}^{N}$ with dimension $D = N - {\rm rank}({\vec \Phi})$. All solutions can be written as
\be
   \vec B_r = \vec B_r^{\rm (sp)} + \vec W \vec y,
\ee
where the columns of the matrix $\vec W \in \mathbb{R}^{N \times D}$ form a basis for the kernel (nullspace) of ${\vec \Phi}$ and $\vec y \in \mathbb{R}^{D}$ is a vector of parameters. $\vec B_r^{\rm (sp)}$ is a specific solution to equation (\ref{eq:PhiBP1}); in the implementation it is obtained using the {\tt MATLAB} function {\tt mldivide}. We now search for the vector $\vec y$ such that the 1-norm (\ref{eq:Br-1norm}) assumes its minimum. This optimization problem is rewritten as a linear program \cite{Boyd04}
\be
   \min_{\vec{s}} \vec 1^T \vec s 
   \quad \mbox{such that} \quad
   \begin{array}{l}
   \vec B_r^{\rm (sp)}  + \vec W \vec y \preceq \vec s    \\
   \vec B_r^{\rm (sp)}  + \vec W \vec y \succeq  -\vec s      \\
   \end{array},
\ee
where $\vec s \in \mathbb{R}^N$ is an auxiliary variable, $\vec 1 \in \mathbb{R}^N$ is a vector of ones and $\preceq$ and $\succeq$ denotes entry-wise comparison. In the standard form of many commercial solvers the optimization problem reads
\be
    \min_{\vec x} \vec f^T \vec x 
    \quad \mbox{such that} \quad
    \vec A \vec x \preceq \vec b.
    \label{eq:linprog}
\ee 
using the further auxiliary variables
\begin{align}
    \vec x & = \begin{pmatrix} \vec y \\ \vec s \end{pmatrix}, \quad
    \vec A    = \begin{pmatrix} \vec W & - \eye \\ - \vec W & -\eye \end{pmatrix},  \nn \\  
    \vec b & = \begin{pmatrix} \vec B_r^{\rm (sp)}  \\ -  \vec B_r^{\rm (sp)}  \end{pmatrix}, \quad 
    \vec f = \begin{pmatrix} \vec 0 \\ \vec 1 \end{pmatrix}, \nn
\end{align}
where $\eye \in \mathbb{R}^{N\times N}$ is the identity matrix and $\vec 0 \in \mathbb{R}^D$ is a vector of zeros.

\begin{figure}[tb]
\centering
\includegraphics[width=\columnwidth]{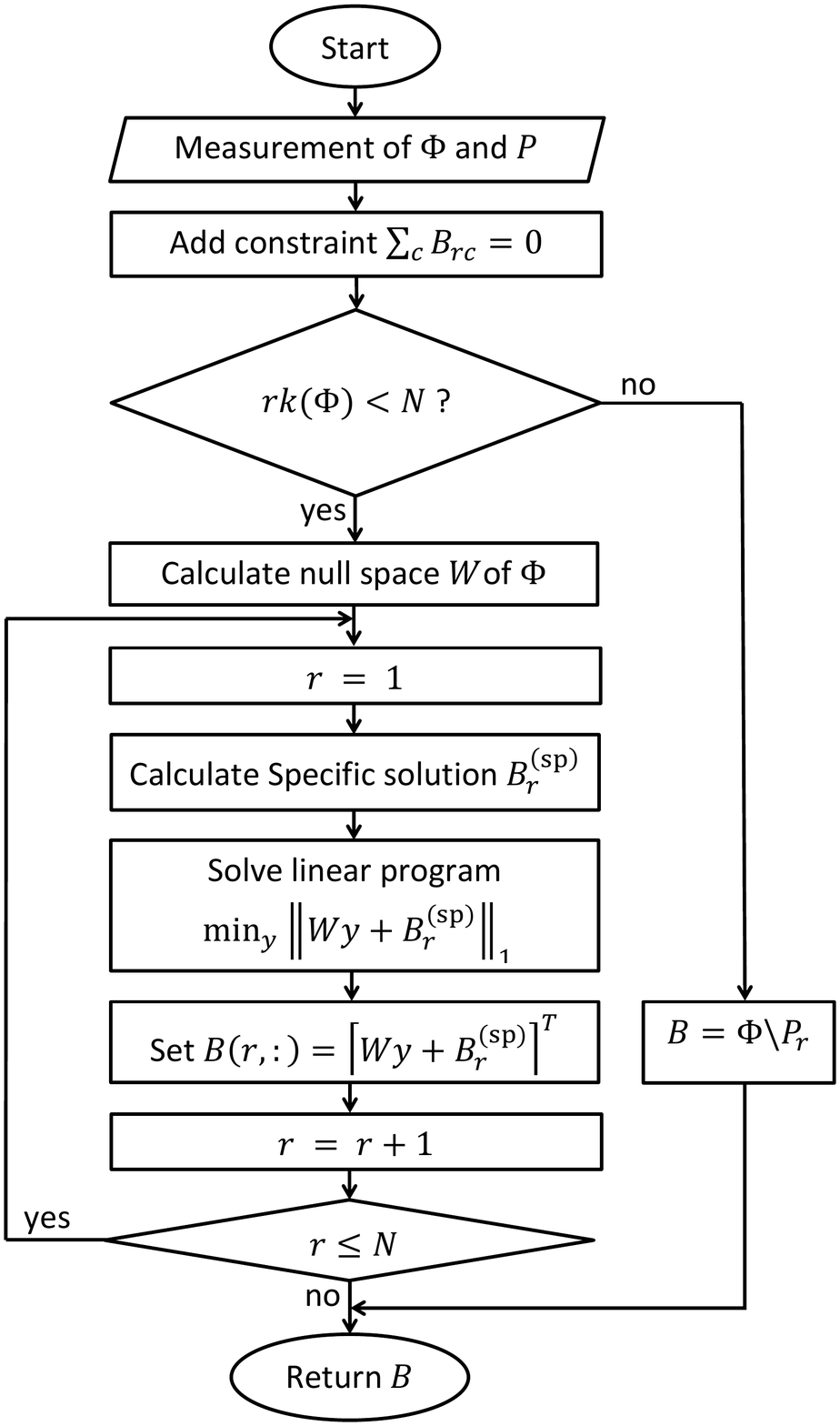}
\caption{
\label{fig:flowchart1}
Flow chart for row-wise reconstruction. {\tt MATLAB} code for the central part is listed in Fig.~\ref{fig:code-row}.
}
\end{figure}

\subsection{Row-Wise Reconstruction}

\begin{figure}[tb]
\begin{Verbatim}[frame=single]
% Case 1: Overdetermined system: 
if rank(Phi) >= N
    Breconstruct = Phi \ Pt;
    
% Case 2: Underdetermined system:    
else
    
    % Calculate nullspace of Phi
    W = null(Phi);
    D = size(W,2);
        
    % treat all rows separately:
    for r=1:N
        
        % one specific solution
        Br_sp = Phi \ Pt(:,r);
                
        % Choose sparse solution
        f = [zeros(D,1); ones(N,1)];
        A = [W, -eye(N); - W, -eye(N)];
        b = [-Br_sp; +Br_sp];       
        y = linprog(f,A,b);
        y = y(1:D);

        % Write result to matrix B 
        Breconstruct(r,:) = (W*y+Br_sp)';
          
    end   
end
\end{Verbatim}
\caption{\texttt{MATLAB} code for row-wise reconstruction}
\label{fig:code-row}
\end{figure}

We first describe the implementation of the row-wise reconstruction of $\vec B$ from the measurement data without any prior knowledge about the entries of $\vec B$, whose basic idea has been outlined in section \ref{sec:idea-rowwise}. The algorithm to solve this problem is illustrated by the flow-chart in Fig.~\ref{fig:flowchart1}. We implement this algorithm in Matlab, the central part of the program code is listed in Fig.~\ref{fig:code-row}.

First data is collected and the matrix $\vec \Phi$ and the vectors $\vec \P_r$ are formed. Then the problem is solved depending on  the rank of the matrix ${\vec \Phi}$. If the system is (over)determined, we directly solve it for $\vec B$ using the Matlab-function \texttt{mldivide}. If the system is underdetermined we proceed row-by-row and reconstruct $\vec B_r$ as described in section \ref{sec:1norm} using the function \texttt{null} to compute the matrix $\vec W$ and the function \texttt{linprog} to solve the linear program (\ref{eq:linprog}).

%

\subsection{Iterative reconstruction}

\begin{figure}[tb]
\centering
\includegraphics[width=\columnwidth]{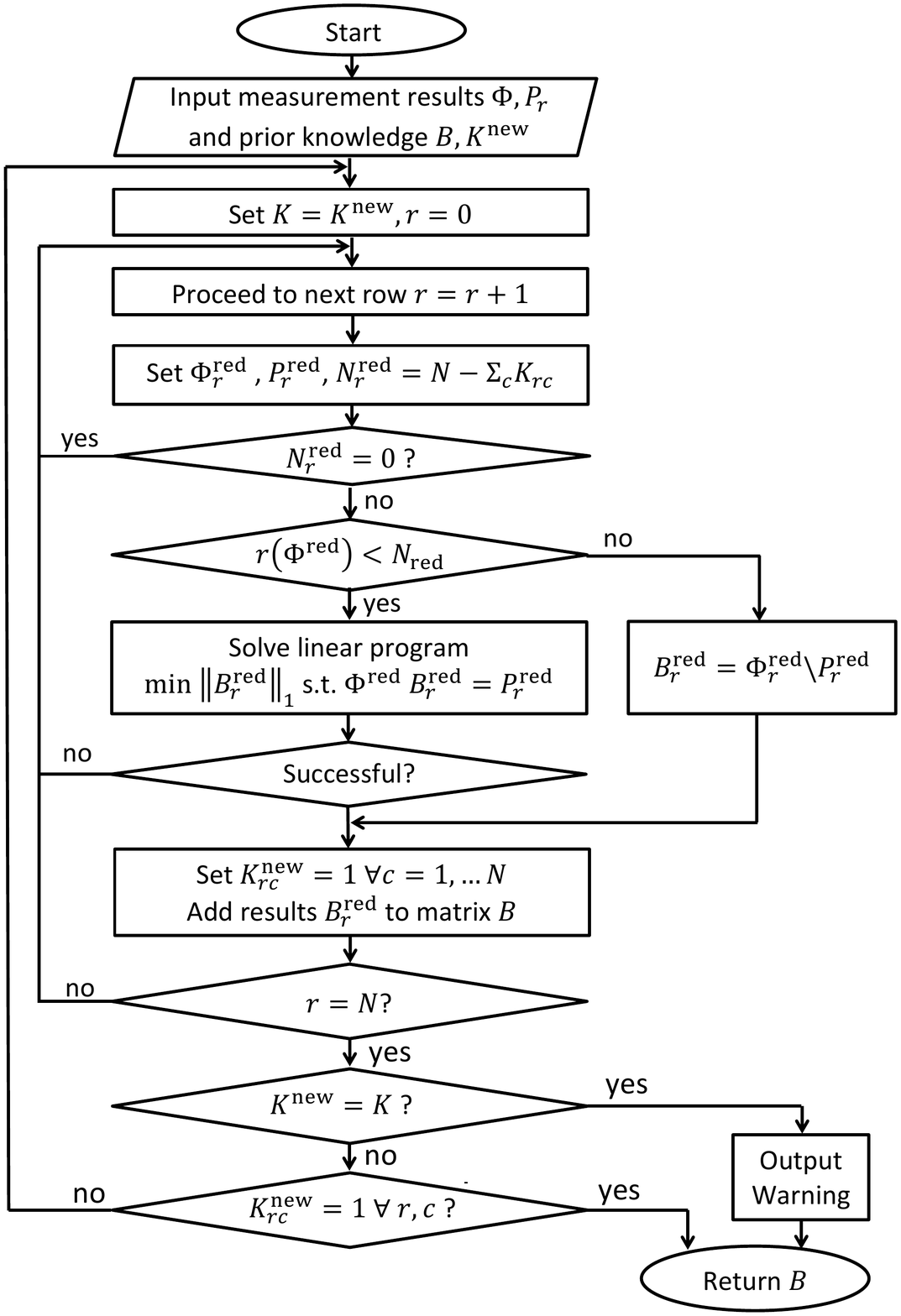}
\caption{
\label{fig:flowchart2}
Flow chart for the iterative reconstruction algorithm. 
}
\end{figure}

In practice we typically have some prior knowledge about the entries of the matrix $\vec B$ as outlined in section \ref{sec:idea-partial}. An algorithm to perform the reconstruction exploiting the prior knowledge is illustrated by the flow-chart in Fig.~\ref{fig:flowchart2}

Most interestingly, we can also use this algorithm to greatly improve the convergence of the reconstruction algorithm. For intermediate values of $M$ we typically face the situation that some rows are successfully reconstructed while others are not. If we have not reconstructed $\vec B$ successfully, but gained further knowledge, we may use it in the following. Having successfully reconstructed the $r$th row of $\vec B$, we also know the entries $B_ {r',r} = B_{r,r'}$ for all other rows $r'$ due to the symmetry of the matrix. We thus propose to perform the reconstruction iteratively. In each step of the reconstruction process we use the initial knowledge about the entries of the matrix $\vec B$ and also the knowledge gained in previous steps. 

The algorithm shown in Fig.~\ref{fig:flowchart2} starts with the input of the measurement results and the prior knowledge of the system which is encoded in the matrices $\vec B$ and $\vec K$. It solves the reconstruction problem iteratively, processing row-by-row in each iteration. For each row, we first calculate the reduced matrices $\vec \Phi_r^{\rm red}$ and $\vec \P_r^{\rm red}$ as defined in section \ref{sec:idea-partial}. Then we attempt to reconstruct the vector $\vec B_r^{\rm red}$ as described above: If the problem is (over)determined we solve the linear system of equations using the Matlab-function \texttt{mldivide}, if it is underdetermined we minimize the 1-norm $\| \vec B_r^{\rm red} \|_1$ subject to the constraint (\ref{eq:PhiBR-red}). If the reconstruction was successful, we add the results to the matrix $\vec B$ and update the knowledge matrix $\vec K$. To facilitate the bookkeeping we define two matrices $\vec K$ and $\vec K^{\rm new}$: We set $\vec K = \vec K^{\rm new}$ at the beginning of each iteration and only modify $\vec K^{\rm new}$ during the step. The iteration stops when the reconstruction has been successfully completed, i.e. $K^{\rm new}_{rc} = 1$ for all $r,c = 1\ldots,N$, or when no further progress has been made, i.e. $K^{\rm new}$ has not been modified during the last step.

One unsolved problem remains: If we do not know the matrix $\vec B$ a priori, how can we know if the reconstruction of the $r$th row has been successful? In the underdetermined case we cannot decide whether the reconstructed values $B_{rc}$ are definitely correct -- but we can decide if they are reasonable in terms of the connectivity of the grid. Power grids are generally very sparse: A single substation is connected to only few other substations. Hence we seek for a solution which is most sparse, i.e. which minimizes
\be
    \| \vec B_r  \|_0 = \mbox{number of non-zero entries of} \; \vec B_r .
\ee
However, we cannot minimize $\| \vec B_r  \|_0$ directly as this is a computationally hard problem. The ingenious contribution of Candes, Romberg, Tao and Donoho \cite{Dono06,Cand06} was the proof that the most sparse solution can be obtained by minimizing $\| \vec B_r  \|_1$ with high probability if a sufficient amount of measurement data is available. Otherwise we find a solution which minimizes the 1-norm, but has a large number of non-zero entries $\| \vec B_r  \|_0$. In algorithm in Fig.~\ref{fig:flowchart2} we thus adopt the definition that the reconstruction of the $r$th row is assumed to be successful if  
(a) the linear system is overdetermined or
(b) the reconstructed row $\vec B_r$ is sufficiently sparse, i.e. $\| \vec B_r  \|_0 \le d_{\rm max}$ for some upper limit $d_{\rm max}$. A typical value used in the following is $d_{\rm max} =15$.

\section{Applications and Performance}

\subsection{Reconstruction from time series}

\begin{figure}[tb]
\centering
\includegraphics[width=\columnwidth]{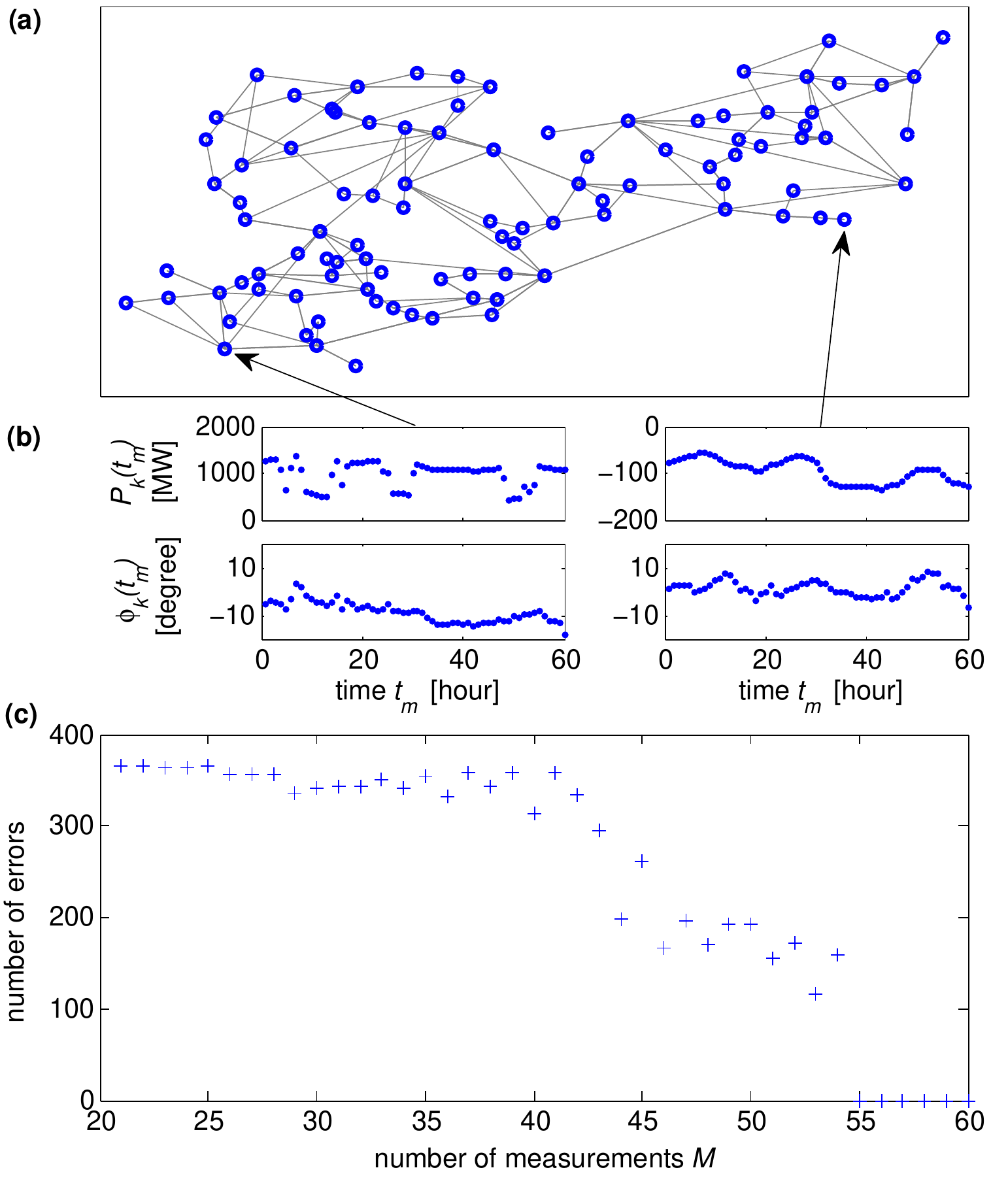}
\caption{
\label{fig:example1}
Example of the complete grid reconstruction for a test grid with $N=117$ nodes.
(a) Test grid from \cite{Barr15}. It is assumed that the network topology, i.e. the location and parameters of the transmission lines (grey) are completely unknown. 
(b) The reconstruction is based on the time series of the nodal power injections $P_k(t)$ and voltage angles $\phi_k(t)$. Two time series are shown. 
(c) Performance of the iterative algorithm shown in Fig.~\ref{fig:flowchart2}. Perfect reconstruction is obtained with $M \ge 55$ measurements.
We use the test grid and the load and generation time series of the scenario 2013 from \cite{Barr15}. For the sake of simplicity we have removed the import/export nodes and rescaled the generation to exactly match the load.
}
\end{figure}

We demonstrate the applicability of network reconstruction from time series measurements for a test grid taken from \cite{Barr15} illustrated in Figure \ref{fig:example1} (a). This data set includes hourly data for demand and generation $P_k(t)$ for one year and all $N=117$ nodes. The nodal voltage angles $\phi_k(t)$ are then obtained by solving the DC approximation (\ref{eq:dcapprox}) for all time steps. The resulting time series of the power injections and angles are shown in Figure \ref{fig:example1} (b) for two nodes as an example.

The algorithm shown in Figure \ref{fig:flowchart2} can now reconstruct the entire network topology - i.e.~all entries of the nodal susceptance matrix $\vec B$ - from the time series data. To evaluate the performance of the algorithm we vary the amount of input data. For each value of $M$ we run the algorithm and count the number of successfully reconstructed entries and the number of reconstruction errors. To account for small numerical errors, we say that an entry $(r,k)$ of the nodal susceptance matrix is successfully reconstructed if
\be
    | B_{r,k}^{\rm reconstructed} - B_{r,k}^{\rm true} | < \epsilon,
\ee   
where the numerical tolerance is chosen as $\epsilon = 10^{-3} \times \max_{ij} |B_{ij}|$.

Figure \ref{fig:example1} (c) shows that $M_{\rm min}=55$ measurements are sufficient to faithfully reconstruct the entire nodal susceptance matrix $\vec B$. Most importantly, a faithful reconstruction is possible in the strongly underdetermined case: The minimum number of measurements $M_{\rm min}=55$ is less than half of the matrix dimension $N=117$. This is possible because we can exploit the sparsity of the matrix $\vec B$ as an additional structural information.

\subsection{Minimum Measurement Requirements and Scaling}

The algorithms presented in figures \ref{fig:flowchart1} and \ref{fig:flowchart2} allow for a reconstruction of the network topology also in the underdetermined case. But how many measurements $M_{\rm min}$ are needed for a faithful reconstruction and how does this number scale with the grid size $N$? 

\begin{figure}[tb]
\centering
\includegraphics[width=\columnwidth]{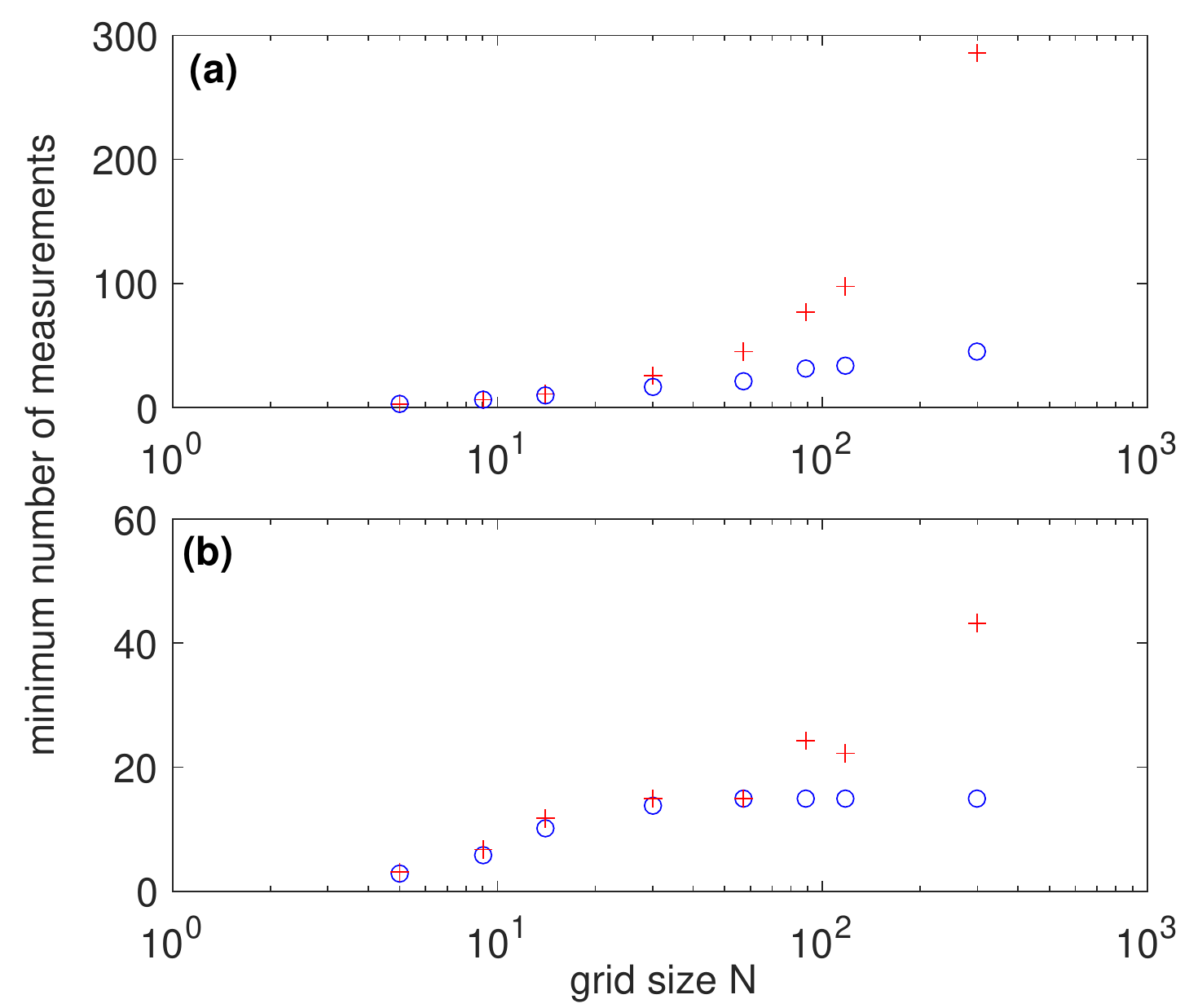}
\caption{
\label{fig:perf-row-case118}
Performance of (a) the row-wise reconstruction algorithm shown in Fig.~\ref{fig:flowchart1} and (b) the iterative algorithm shown in Fig.~\ref{fig:flowchart2}. We have tested the algorithms for several test grids from \cite{matpower,Flis13} for synthetic power injection data, choosing either the $\phi_k$  (blue circles) or the $P_k$ (red crosses) randomly (see text for details). Shown is the minimum number of measurements $M$ needed to obtain a faithful reconstruction as a function of the grid size $N$. Results have been averaged over 10 random realizations.
}
\end{figure}

To systematically study the efficiency of the reconstruction algorithms, we consider various test grids of different size taken from \cite{matpower,Flis13} and use randomized synthetic data for the time series. We consider two different types of time series data. First we draw the voltage phase angles $\phi_k(t)$ uniformly at random from the interval $[-\pi/8,+\pi/8]$ (type I). This scenario is not realistic but close to the original mathematical work \cite{Dono06,Cand06}, where random Gaussian sampling vectors are considered. Second, we choose the power injections $P_k(t)$ at random from a normal distribution with mean zero and standard deviation as in the original test grid (type II). For both types and both reconstruction algorithms (row-wise and iterative), we run the reconstruction algorithm as a function of the number of measurements $M$ taken into account and determine the minimum number of measurements $M_{\rm min}$ for a faithful reconstruction, i.e. zero reconstruction errors. Each numerical experiment is repeated 10 times.

First, we observe that the iterative algorithm allows for a faithful reconstruction from underdetermined input data for all grids under consideration and both data types. The minimum number of measurements $M_{\rm min}$ is much smaller than the grid size $N$ in all cases. The row-wise algorithm performs less well as it does not exploit the symmetry of the matrix $\vec B$.
 
Second, efficient reconstruction is possible with both algorithms for the type I input data. The theory of compressed sensing shows that a reconstruction of sparse vectors is possible with high probability from a number of measurements scaling only logarithmically with the problem dimension $N$ if some conditions are satisfied  \cite{Dono06,Cand06}. This logarithmic scaling is well confirmed by the results of our numerical experiments. For the iterative algorithm the scaling seems to be even slower.
 
However, the proofs for the efficiency of compressed sensing depend on two features of the sampling vectors: isotropy and incoherence. Roughly speaking, isotropy means that all directions in the vector space are sampled equally well. Incoherence guarantees that the redundancy in the information gained by two measurements is small. Both features are no longer guaranteed for type II input data. In fact, we observe that the row-wise reconstruction algorithm does not perform well in this case. The minimum number of measurements $M_{\rm min}$ is only slightly smaller than the problem dimension $N$ such that the iterative algorithm performs much better.

\subsection{Reconstruction from flow measurements}

\begin{figure}[tb]
\centering
\includegraphics[width=0.48\columnwidth]{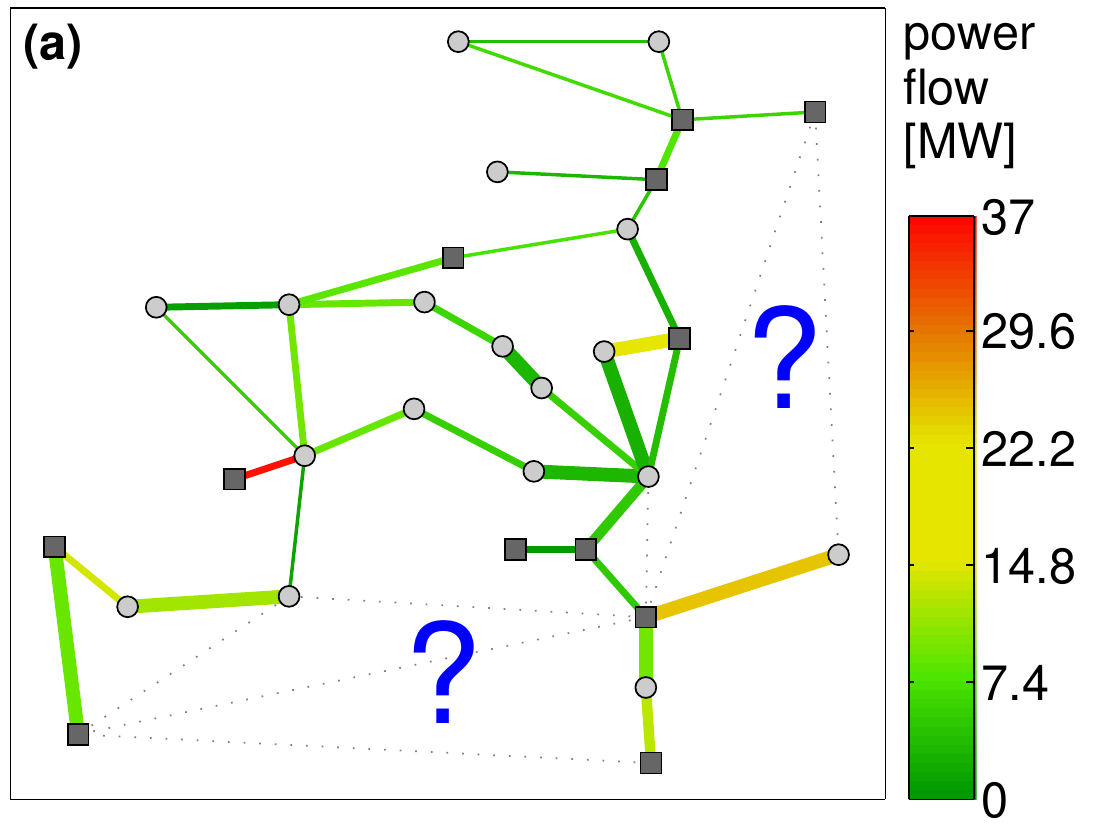}
\includegraphics[width=0.48\columnwidth]{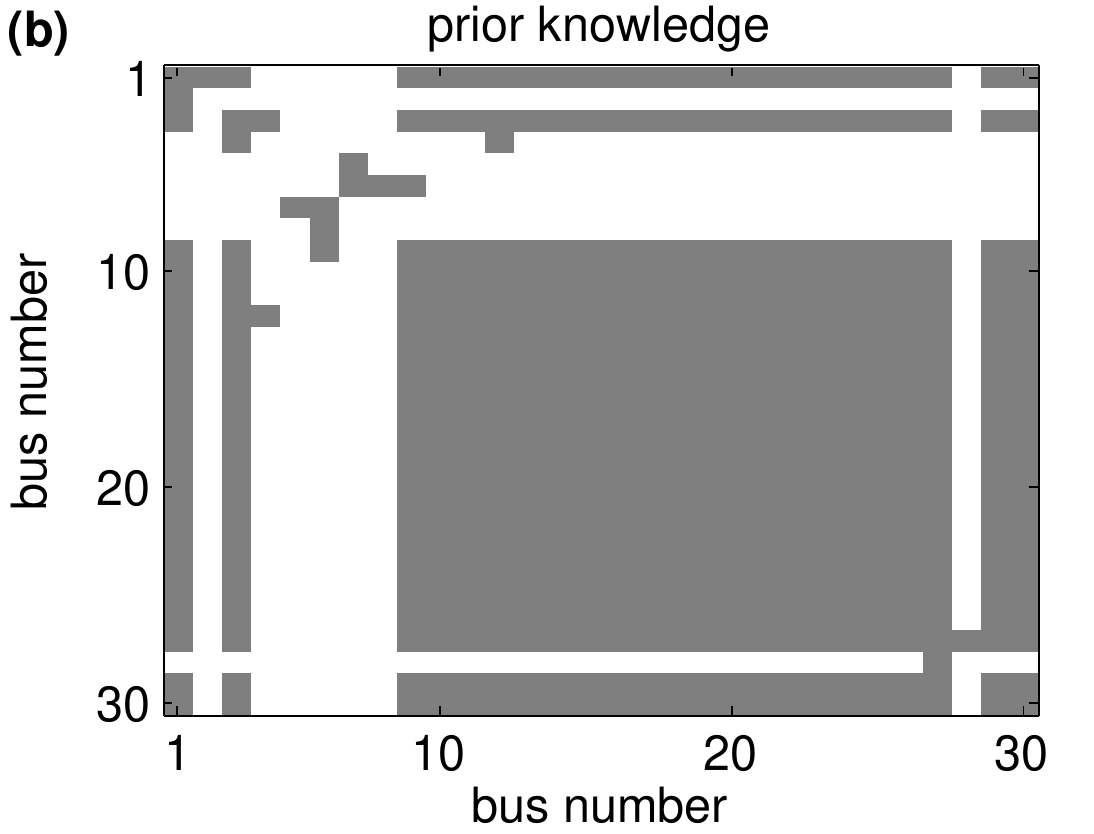}\\[4mm]
\includegraphics[width=0.48\columnwidth]{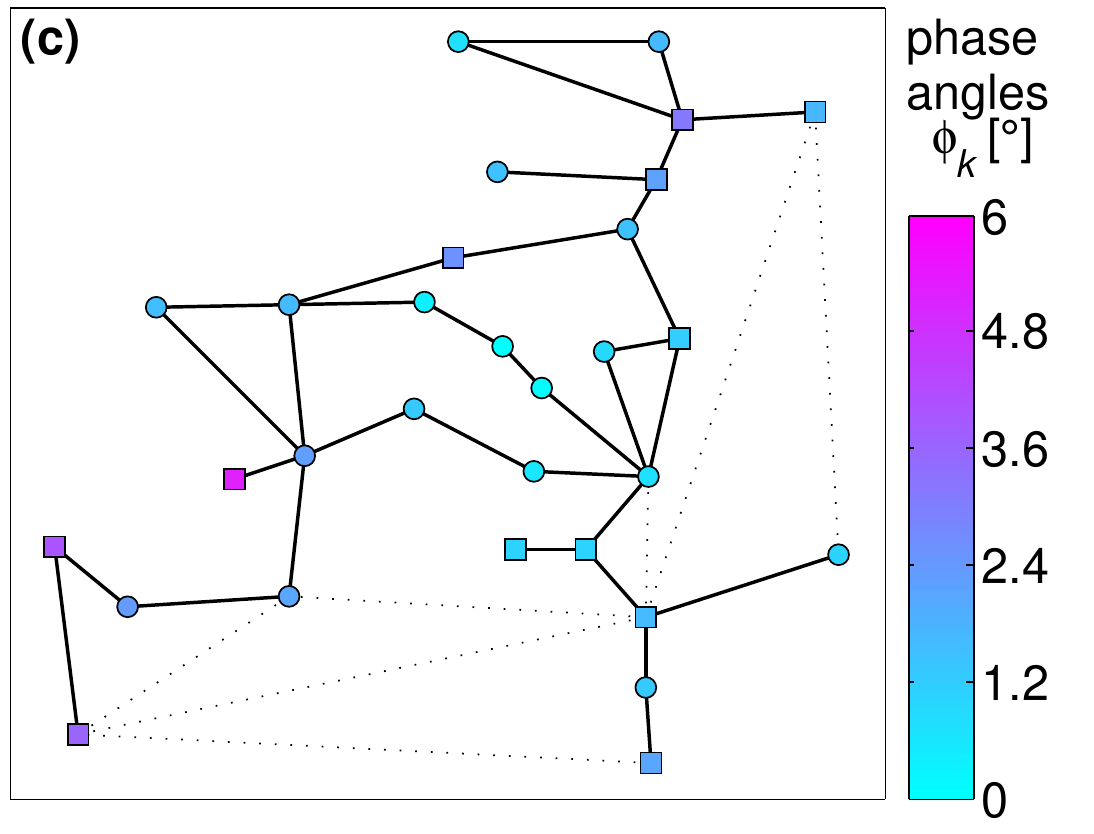}
\includegraphics[width=0.48\columnwidth]{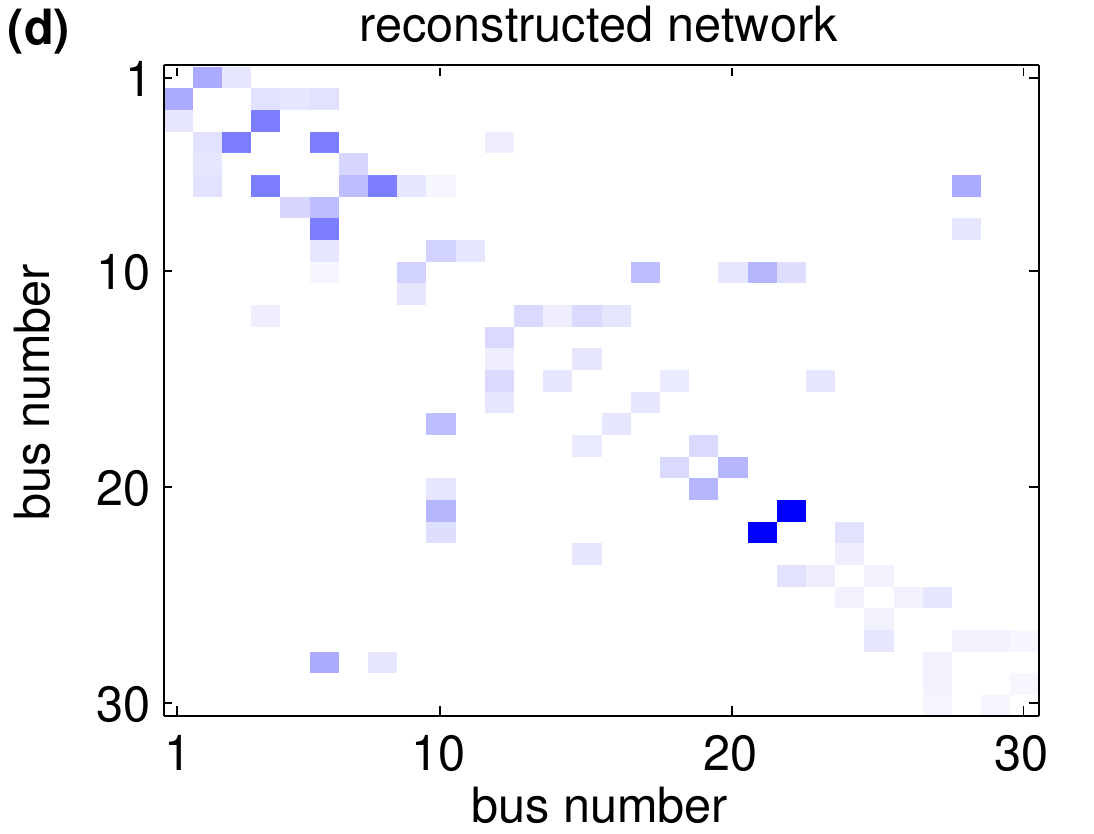}
\caption{
\label{fig:flow1}
Reconstruction of missing information from flow measurements
(a) The IEEE 30-bus test grid. Real power flows are measured on the colored lines. In two regions the grid topology is assumed to be unknown.
(b) The prior knowledge matrix $K$. Grey shading indicates that the respective line parameter
$B_{jk}$ is known. The white region is completely unknown.
(c) The voltage phase angles $\phi_k$ are reconstructed from the power flow measurements using standard state estimation methods.
(d) The completely grid topology is perfectly reconstructed using the algorithm shown in Fig.~\ref{fig:flowchart2}. 
}
\end{figure}

The iterative algorithm allows to take into account prior knowledge to efficiently reconstruct the missing information about the grid topology. Such a problem can arise in practice when a larger part of the grid is subject to damages or attacks. An example of such a situation is depicted in figure \ref{fig:flow1} (a). It is assumed that we have no reliable information about the grid in two areas indicated by the question marks. However, we do have information about the connections of all remaining nodes and we assume that we can measure the real power flow along the colored solid lines. The prior knowledge of the grid topology is encoded in the matrix $\vec K$, which is illustrated in figure \ref{fig:flow1} (b). White entries indicate where we do not know the entries of the matrix $\vec B$.

The missing information is found by combining classical state estimation and network reconstruction. First, the voltage phase angles for all nodes of the network are reconstructed as described in section \ref{sec:stateest}, the results being shown in figure \ref{fig:flow1} (c). Second, the iterative reconstruction algorithm shown in figure \ref{fig:flowchart2} is applied using the measured power injections, the prior knowledge $\vec B$ and $\vec K$ as well as the estimated voltage phase angles as input data. The algorithm then faithfully reconstructs all the missing information about the grid topology. The reconstructed values of the line susceptances $X_{sr}^{-1}$
shown in figure \ref{fig:flow1} (d) exactly match the true values. 
Most importantly, the full reconstruction is possible already for $M=1$ in this case, i.e. with a single snapshot of the power injections $\vec P$, making use of the prior knowledge.

\section{Conclusion and Outlook}

We have introduced two algorithms to reconstruct the structure of a power grid from nodal measurements only. One algorithm is straightforward and row-based, reconstructing the lines of the grid node by node. The second is iterative and takes into account both prior knowledge about the presence of absence of lines as well as the knowlegde generated by that algorithm during previous steps. We have demonstrated how these algorithms can be used to reconstruct the entire network structure from time series or missing information on the grid topology from a single snapshot. 

The presented algorithms exploit several structural properties of power grids to reduce the number of necessary measurements. Power grids are typically very sparse, i.e. each substation is connected to only few other substations. Hence, methods from compressed sensing can be used which allow for a faithful reconstruction also in the underdetermined case. The iterative reconstruction algorithm makes use of the symmetry of the nodal susceptance matrix. Loosely speaking the algorithm solves the simple parts of the reconstruction problem first and than uses the gained information for the remaining parts. We have shown that this trick leads to a vast reduction of the measurement resources. 

The present work  presents a step towards hybrid state estimation/network reconstruction algorithms. In future smart grids one can face the situation where both nodal variables and the grid topology are subject to uncertainties or lacking information. A pure state estimation starting from perfect knowledge of the network structure is then no longer sufficient. Advanced grid monitoring algorithms must be able of coping with missing information in both state and structure.

\section*{Acknowledgements}

We thank D.~Gross and R.~Kueng for inspiring discussions.
We gratefully acknowledge support from the Helmholtz Association 
(via the joint initiative ``Energy System 2050 - A Contribution of the Research Field Energy''
and the grant no.~VH-NG-1025 to D.W.) and
the Federal Ministry of Education and 
Research (BMBF grant nos.~03SF0472B and ~03SF0472E to D.W. and M.T.).




%

\begin{IEEEbiographynophoto}{Farnaz Basiri}
Farnaz Basiri received the B.Sc. and M.Sc. degree in electrical power engineering from the RWTH Aachen University, Aachen, Germany in 2013 and 2015, respectively. Since 2015 she joined the Reaserch Group at Reaserch Center J\"ulich, Germany as a Research Associate and is currently pursuing the Ph.D. degree.
\end{IEEEbiographynophoto}

\begin{IEEEbiographynophoto}{Jose Casadiego}
received his Licentiate degree in Physics from the University of Carabobo (Venezuela) in 2009. He completed his PhD studies in the International Max Planck Research School for Physics of Biological and Complex systems (G\"ottingen) in 2016 working on model-free approaches for revealing interactions in complex networks from collective dynamics. He now works as a Postdoctoral Researcher at the Network Dynamics Group of the Max Planck Institute for Dynamics and Self-Organization.
\end{IEEEbiographynophoto}

\begin{IEEEbiographynophoto}{Marc Timme}
    studied physics and applied mathematics at the Universities of W\"urzburg (Germany), Stony Brook (New York, USA) and G\"ottingen (Germany). He holds an M.A. in Physics from Stony Brook and a Doctorate in Theoretical Physics (G\"ottingen). After postdoctoral and visiting stays at Cornell University (New York, USA) and the National Research Center of Italy (Sesto Fiorentino) he is heading the Max Planck Research Group on Network Dynamics at the Max Planck Institute for Dynamics and Self-Organization. He is Adjunct Professor at the University of G\"ottingen. 
\end{IEEEbiographynophoto}

\begin{IEEEbiographynophoto}{Dirk Witthaut}
    received his Diploma (MSc) and PhD in Physics from the Technical University of 
    Kaiserslautern, Germany, in 2004 and 2007, respectively. He has been a Postodoctoral 
    Researcher at the Niels Bohr Institute in Copenhagen, Denmark and at the Max Planck 
    Institute for Dynamics and Self-Organization in G\"ottingen, Germany and a
    Guest Lecturer at the Kigali Institue for Science and Technology in Rwanda.
    Since 2014 he is heading a Research Group at Forschungszentrum J\"ulich, Germany and he is
    Junior Professor at the University of Cologne, Germany.   
\end{IEEEbiographynophoto}

\end{document}